# Recent Progress at SLAC Extracting High Charge from Highly-Polarized Photocathodes for Future-Collider Applications[*]

J. E. Clendenin[a,†], A. Brachmann[a], E. L. Garwin[a], S. Harvey[a], J. Jiang[a],
R. E. Kirby[a], D. -A. Luh[a], T. Maruyama[a], R. Prepost[b], C. Y. Prescott[a],
J. L. Turner[a]

[a] *Stanford Linear Accelerator Center, 2575 Sand Hill Road, Menlo Park, CA 94025, USA*
[b] *Department of Physics, University of Wisconsin, Madison, WI 53706, USA*

## Abstract

Future colliders such as NLC and JLC will require a highly-polarized macropulse with charge that is more than an order of magnitude beyond that which could be produced for the SLC. The maximum charge from the SLC uniformly-doped GaAs photocathode was limited by the surface charge limit (SCL). The SCL effect can be overcome by using an extremely high ($\geq 10^{19}$ cm$^{-3}$) surface dopant concentration. When combined with a medium dopant concentration in the majority of the active layer (to avoid depolarization), the surface concentration has been found to degrade during normal heat cleaning (1 hour at 600 °C). The Be dopant as typically used in an MBE-grown superlattice cathode is especially susceptible to this effect compared to Zn or C dopant. Some relief can be found by lowering the cleaning temperature, but the long-term general solution appears to be atomic hydrogen cleaning.

PACS : 29.25.Bx, 73.50.Pz, 73.61.Ey, 85.60.Ha

Keywords: Photocathodes, photoemission, semiconductors, surface charge limit

Contributed to the *10th Workshop on Polarized Sources and Targets,*
September 22-26, 2003, Novosibirsk, Russia

---

[*] Work supported by Department of Energy contract DE-AC03-76SF00515.
[†] Corresponding author. Tel.: +1-650-926-2962; fax: +1-650-926-8533.
*E-mail address*: clen@slac.stanford.edu (J.E. Clendenin)

## Introduction

The surface charge limit (SCL), which affects negative electron affinity (NEA) semiconductor photocathodes when one attempts to extract very high current densities using a laser tuned to the band gap energy, was first observed in 1978 with the first SLAC GaAs source, then more completely characterized in 1991 during commissioning of the SLC polarized electron source [1]. The phenomenon has been described in the literature [2]. An example of a severe SCL effect on emission from a GaAs-type photocathode is shown in Fig. 1. For the figure, the laser spatial and temporal profiles were held constant while the laser energy at the wavelength corresponding to the band gap was increased in approximately equal steps. The suppression of emission after the initial few nanoseconds develops at relatively low laser energy. The suppressed emission stabilizes after about 200 ns.

The SLC bunch train consisted of up to $8\times10^{10}$ e$^-$ (at the source) in each of 2 microbunches separated by 60 ns. At this level the bunches were already affected by the SCL, with the effect increasing with time as the quantum efficiency (QE) decreased. The JLC/NLC requires 192 microbunches each separated by 1.4 ns. Each microbunch is required to have $0.75\times10^{10}$ e$^-$ at the interaction point (IP). It is assumed here that the source must produce $1.5\times10^{10}$ e$^-$ or a total of $2.9\times10^{12}$ e$^-$ in the 270 ns bunch train, which is an order of magnitude more charge than in the SLC train, which means the "standard"

SLC cathodes (MOCVD-grown[1] 100-nm strained-layer GaAsP/GaAs uniformly Zn-doped at $5\times10^{18}$ cm$^{-3}$) will not work for JLC/NLC.

## Suppression of the Surface Charge Limit

Very early it was observed that at least 3 factors–the QE (relative to the initial QE), the excitation wavelength, and the dopant concentration–contribute to the effect of the SCL on GaAs-type photocathodes [3]. In 1998 the Nagoya group demonstrated that with a superlattice having a high ($4\times10^{19}$ cm$^{-3}$) dopant concentration at the surface, a train of 12-ns wide microbunches could be produced with intensities up to the space charge limit of the 70-kV gun (20 nC or 1.65 A) with no SCL effect observed. This performance was attributed to a combination of the high dopant concentration and the large band gap of the superlattice structure. The effectiveness of the high dopant concentration was confirmed at SLAC by preparing a series of unstrained 100-nm GaAs cathodes each with uniform doping [2]. As shown in Fig. 2, it appears that a dopant concentration $\geq 2\times10^{19}$ cm$^{-3}$ is sufficient for a modestly well-activated surface. A high dopant concentration decreases the polarization of the emitted electrons, but fortunately only the final few nanometers at the surface need be highly doped. The technique of "gradient doping" (low concentration in the bulk, high at the surface) is now routinely used for providing polarized beams for the SLAC linac.

The "standard" SLC cathode was modified with gradient doping [4] and used in the first phase (2002) of a parity violating experiment (E-158) at SLAC [5]. The experiment required a pulse length of ~300 ns, similar to the JLC/NLC macropulse requirement. The flashlamp-pumped Ti:sapphire laser [6] for the source could produce up to 200 μJ in this pulse length. Using this cathode, a total charge of $2.3\times10^{12}$ e⁻–nearly equal to the total charge required for the JLC/NLC macropulse–was produced at the source in a 100-ns pulse using a 20-mm diameter laser spot at the cathode, and $1.4\times10^{12}$ e- for a 14-mm spot–corresponding to 3.7 and 2.2 A respectively [4].

---

[1]Bandwidth Semiconductor (formerly SPIRE Corp.), 25 Sagamore Park Dr., Hudson, NH 03051 USA.

More recently, the SCL properties of GaAsP/GaAs strained superlattice (SL) cathodes–first introduced by the Nagoya group [7]–were explored [8]. The SLAC structures were MBE-grown by SVT Associates[2]. These SL cathodes have a larger band gap than the "standard" SLC cathodes, which generally results in a higher (QE). The higher QE permits the SCL to be probed at higher charge and current densities. In addition, the flashlamp-pumped Ti:sapphire (flash-Ti) laser used for E-158 was successfully Q-switched, which greatly increased the available laser power. The resulting maximum current of 5.5 A, which exceeded the 4.8 A peak current required for JLC/NLC, is shown in Fig. 3. The curvature is due to the space charge limit of the 120-kV gun. This type of cathode was used in the final run (2003) of E-158, producing a polarization of 90% (online value). The flash-Ti laser was not Q-switched during the experiment. NLC requirements and the results achieved at SLAC are summarized in Table 1.

Although the gradient doping technique has been shown to work, in practice it presents a serious problem. The cathodes must be activated by adding Cs and an oxidizer to the atomically-clean surface. The final cleaning is universally done in accelerator sources by heating the cathode to 600 ºC for ~1 hour. Frequently, multiple cleanings are necessary to achieve a high QE. The Zn-doped MOCVD-grown cathodes can be heat cleaned at least twice with no serious diminution of the gradient doping. However, the Be-doped MBE-grown cathodes must be cleaned at temperatures well below 600 ºC to prevent the diffusion of the Be dopant at the surface, which sometimes results in a poor activation as indicated by a low initial QE. One solution is to use atomic hydrogen cleaning (AHC), which does not require temperatures >400 ºC [9]. Ideally an AHC system would be integrated with the gun. An alternative may be to coat the AHC-cleaned crystal with Sb and then transport it to the gun where the remainder of the activation can take place after the Sb is evaporated at a modest temperature.

---

[2]SVT Associates, 7620 Executive Dr., Eden Prairie, MN 55344 USA.

## Conclusion

Gradient doping has been shown to suppress the SCL sufficiently to allow peak currents in excess of NLC/JLC requirements. Implementing for JLC/NLC gradient doping using the best photocathode structures available today will require a low-temperature heat-cleaning technique such as atomic hydrogen cleaning.

## References

[1] M. Woods *et al.*, J. Appl. Phys. 73 (1993) 8531.

[2] G. A. Mulhollan *et al.*, Phys. Lett. A 282 (2001) 309, and references therein.

[3] H. Tang, in Proc. of the Workshop on Photocathodes for Polarized Electron Sources for Accelerators, SLAC-432 Rev. (April 1994), p. 344.

[4] T. Maruyama *et al.*, Nucl. Instrum. and Meth. A 492 (2002) 199.

[5] J. L. Turner *et al.*, Proc. 8th European Part. Acc. Conf. (2002) 1419.

[6] A. Brachmann *et al*., Proc. SPIE Int. Soc. Opt. Eng.4632 (2002) 211.

[7] T. Nishitani *et al*., in SPIN 2000, 14[th] Int. Spin Phys. Symp., AIP Cong. Proc. 570 (2000), p. 1021.

[8] T. Maruyama *et al*., "A Systematic Study of Polarized Electron Emission from Strained GaAsP/GaAs Superlattice Photocathodes," to be submitted to Appl. Phys. Lett.

[9] T. Maruyama *et al*., Appl. Phys. Lett. 82 (2003) 4184.

## Figure Legends

Fig. 1. Electron emission current from a GaAs-type photocathode showing a severe SCL effect. The laser energy was increased in approximately equal steps keeping the spatial and temporal (flat) profiles constant.

Fig. 2. Electron emission current from uniformly doped unstrained, 100-nm GaAs cathodes. The QE for each sample was 0.45, 0.9, 0.4 and 0.4% in the order of increasing dopant concentration. The laser energy was increased in approximately equal steps to 150 W/cm$^2$ keeping the spatial and temporal (flat) profiles constant.

Fig. 3. Electron emission current produced by the flashlamp-pumped Ti:sapphire laser.

**Fig. 1.**

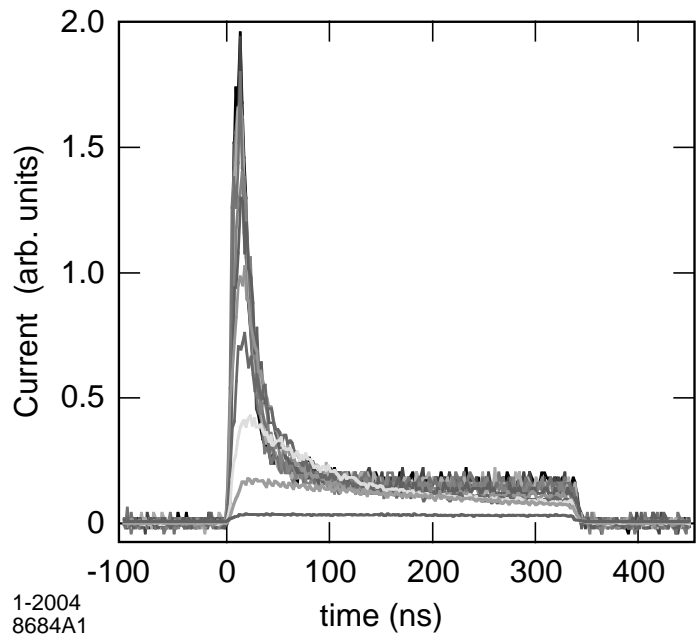

1-2004
8684A1

**Fig. 2**.

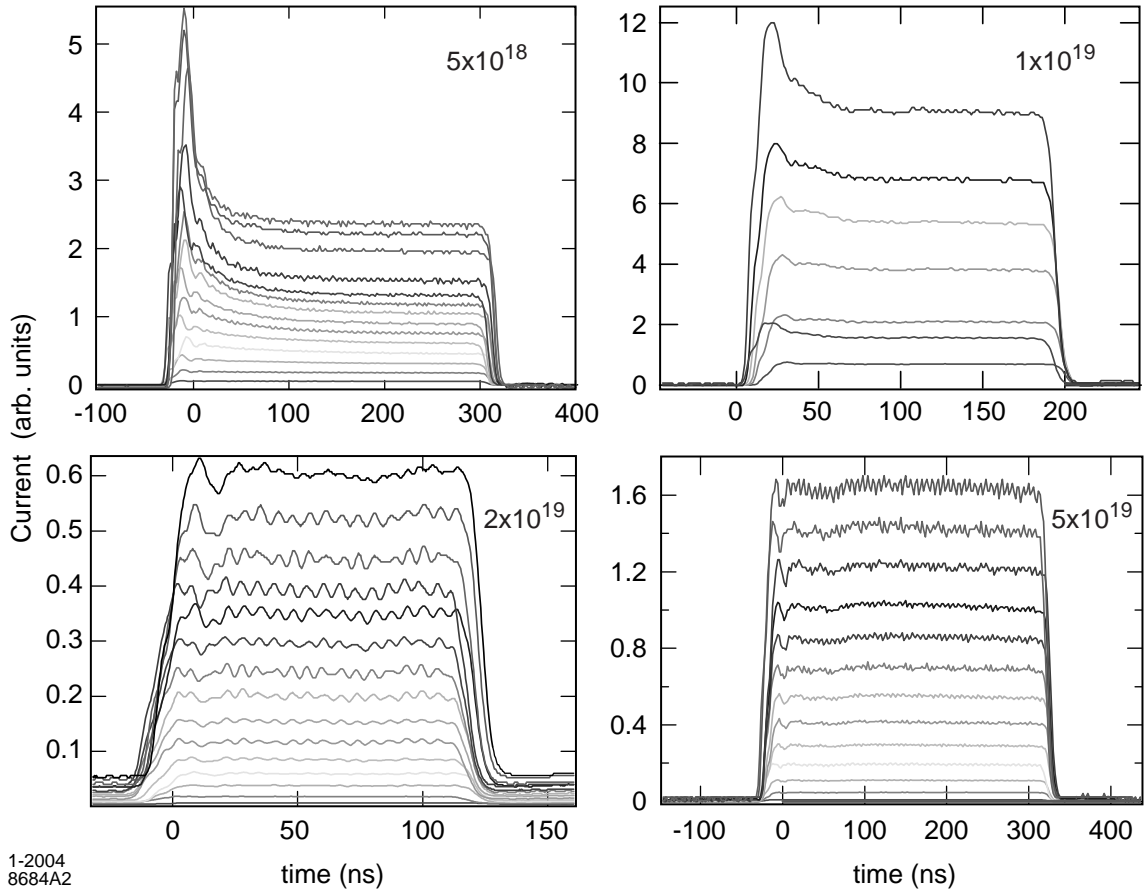

**Fig. 3.**

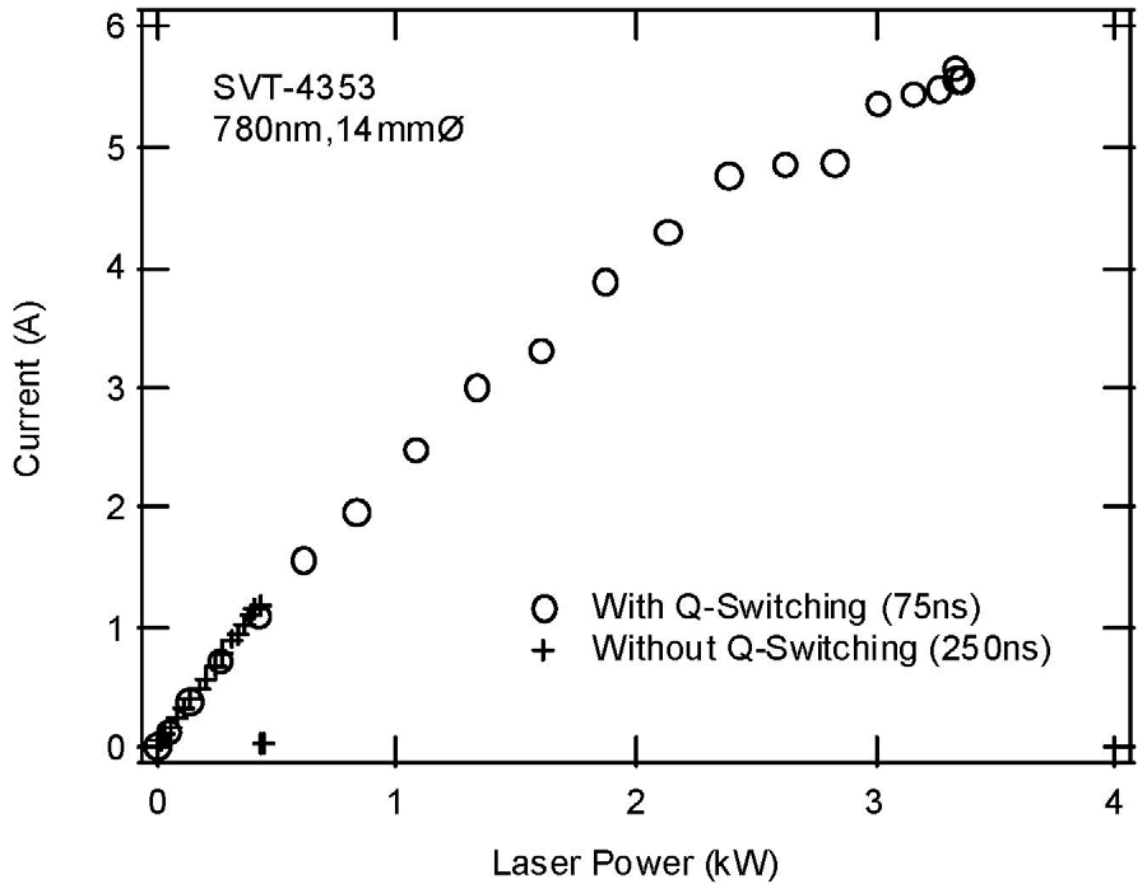

Table 1. Summary of SLAC Results.

| Cathode | Laser Energy μJ | λ nm | Pulse Length ns | Charge per pulse e⁻/pulse | QE % | Laser Pk Power kW | Peak Current A | Dia. mm | Current Density Acm$^{-2}$ |
|---|---|---|---|---|---|---|---|---|---|
| **NLC requirement** | | | | | | | | | |
| micropulse | | | 0.5 | $1.5 \times 10^{10}$ | | | 4.5 | | |
| macropulse | | | 270 | $2.9 \times 10^{12}$ | | | | | |
| **2002** [4] | | | | | | | | | |
| BW-Semi strained-layer | 200 | 805 | 100 | $2.3 \times 10^{12}$ | 0.31 | 2 | 3.7 | 20 | 1.1 |
| GaAsP/GaAs | 150 | 805 | 100 | $1.4 \times 10^{12}$ | 0.25 | 1.5 | 2.2 | 14 | 1.5 |
| **2003** [8] | | | | | | | | | |
| SVT strained GaAsP/GaAs | 247 | 780 | 75 | $2.6 \times 10^{12}$ | 0.44 | 3.3 | 5.5 | 14 | 3.6 |
| superlattice | 225 | 780 | 75 | $2.0 \times 10^{12}$ | 0.42 | 3 | 4.3 | 10 | 5.5 |